\documentclass[twocolumn,secnumarabic,amssymb, amsmath,nobibnotes, aps,prb]{revtex4}

\makeatletter
\def\@dotsep{4.5}
\makeatother

\usepackage{graphicx}
\usepackage{amsmath}
\usepackage{mathrsfs}

\begin{document}
\title{Laser frequency stabilization to excited state transitions using electromagnetically induced transparency in a cascade system}
\author{R. P. Abel, A. K. Mohapatra, M. G. Bason, J. D. Pritchard, K. J. Weatherill, U. Raitzsch and C. S. Adams \footnote[2]{email: c.s.adams@durham.ac.uk} }
\address{Department of Physics, Durham University, Rochester Building, South Road, Durham DH1 3LE, UK}
\date{January 2009}

\begin{abstract}

\noindent  We demonstrate laser frequency stabilization to excited state transitions using cascade electromagnetically induced transparency. Using a room temperature Rb vapor cell as a reference, we stabilize a first diode laser to the $\rm{D_2}$ transition and a second laser to a transition from the intermediate 5P$_{3/2}$ state to a highly excited state with principal quantum number $n=19 - 70$. A combined laser linewidth of 280 $\pm$ 50 kHz over a 100 $\mu$s time period is achieved. This method may be applied generally to any cascade system and allows laser stabilization to an atomic reference in the absence of a direct absorption signal.

\end{abstract}

\maketitle

Many experiments require a stable frequency in the optical region of the electromagnetic spectrum \cite{hall06}. Typically, an optical frequency reference is established by locking a laser to an optical transition from an atomic ground state. Many techniques exist for locking to an atomic reference, including dichroic atomic vapour laser locking (DAVLL) \cite{cher94,corw98,mill07}, combined saturated absorption and DAVLL \cite{harr08,tino03,wasi02}, polarization spectroscopy \cite{pear02}, Sagnac interferometry \cite{robi02,jund03}, frequency modulation (FM) spectroscopy \cite{bjor80} and modulation transfer spectroscopy \cite{shir82,zhan03,mcca08}. Recently, electromagnetically induced transparency (EIT) \cite{harris,flei05} has been used as a dispersive reference to lock the relative frequency of two lasers to an atomic ground-state hyperfine splitting \cite{bell07}. Also Doppler-free two-color spectroscopy has been used to lock to excited-excited state transitions where the lower state is populated  \cite{scho08}. A limitation of these schemes is that one is restricted to wavelengths corresponding to transitions with a significant population in the lower state or with sufficiently large Einstein A-coefficient to produce an absorption signal. Where this is not the case the alternative is to use a frequency comb or a reference cavity stabilized to a ground state transition \cite{bohl06}. However, these methods are limited by the stability of the cavity, which is influenced by environmental factors such as temperature and pressure. 

\begin{figure}[b] 

\centering
\includegraphics[]{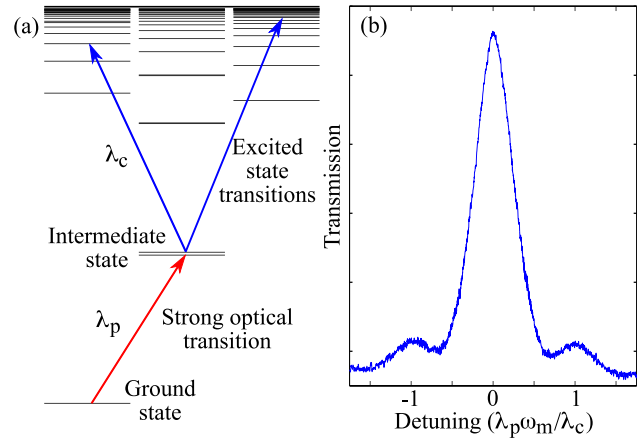}
\caption{(a) Energy level diagram of a two photon transition to a highly excited state in a typical alkali atom system. By probing the strong ground state transition the excited state transition is detected using EIT. (b) A typical EIT spectrum as the coupling laser is scanned 
across resonance. An EIT resonance is oberved at both the carrier and sideband frequencies, $\omega_0\pm(\lambda_p/\lambda_c)\omega_{\rm{m}}$.} 

\label{fig:energy_levels&EIT}

\end{figure}

In this letter, we demonstrate a technique which enables laser stabilization to excited state transitions based on atomic coherences. In contrast to previous laser stabilization schemes, which rely on having a significant population in the lower state our technique allows laser stabilization to excited-excited state transitions with no population in either state.  Also, our scheme uses a probe laser to stabilize a second laser at a completely different wavelength, as shown in Fig~\ref{fig:energy_levels&EIT}~(a). The technique uses EIT to transfer information about a weak excited-excited state transition to a strong probe transition. This allows laser stabilization to transitions with small A-coefficients (of order 100 Hz) such as transitions to Rydberg states. In addition, as the technique is based on an atomic coherence one can obtain resonance widths which are smaller than the natural broadening of the lower state. In principle, the technique is applicable to any excited state transition allowing a large number of additional potential optical frequency references. 

The specific case we consider is  based on cascade Rydberg \cite{gallagher} EIT in a Rubidium vapour cell \cite{moha07, baso08} at room temperature. This application is particularly useful for experiments involving highly excited Rydberg states \cite{weat08, moha08}. We present details of our experimental setup and give examples of locking signals produced at various laser powers and for various Rydberg states. The stability of the lock is demonstrated by results from an experiment on EIT of cold Rydberg ensembles \cite{weat08}.

In order to produce an error signal we use frequency modulation spectroscopy \cite{bjor83}. The probe beam is modulated to give sidebands above and below the EIT resonance. These sidebands then beat with the probe beam to produce a detector signal at the modulation frequency. Using a phase sensitive detection scheme dispersion and absorption components can be recovered to form an error signal. The signal detected is at the modulation frequency, so the bandwidth of the feedback is not limited by the resonance width \cite{hall83}. With the probe beam locked and the coupling beam scanning through resonance an EIT feature is observed  on the transmission of the probe at both the carrier and sideband frequencies as shown in Fig.~\ref{fig:energy_levels&EIT}~(b) . Note that due to the Doppler mismatch between the probe and coupling beams, the frequency offset of the sidebands is scaled by a factor of $\lambda_p / \lambda_c$.   

\begin{figure}[t] 
\centering
\includegraphics[]{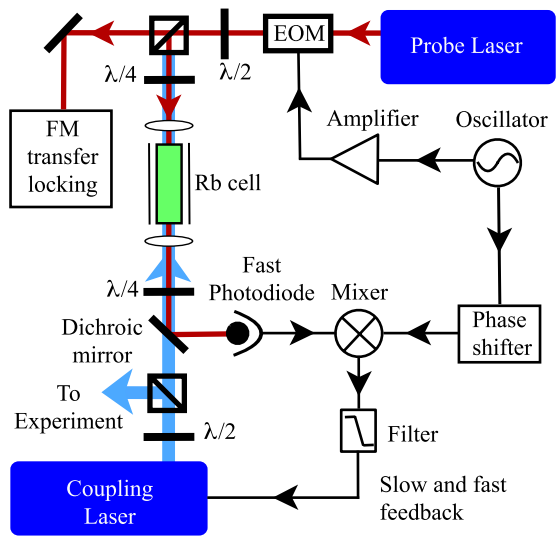}
\caption{Schematic experimental setup for EIT laser stabilization. The coupling and probe laser are counter propagated through a magnetically shielded Rb vapour cell. The probe beam is modulated using an EOM and detected using a fast photodetector. This signal is mixed with the oscillator to produce an error signal that is fed back to the coupling laser.    }
\label{fig:setup}
\end{figure}

In our experiment, the cascade system consists of a weak probe beam resonant with the $^{87}$Rb 5s$^2$S$_{1/2}$ (F=2)   $\rightarrow$ 5p$^2$P$_{3/2}$  (F$^\prime$) transition followed by an intense coupling beam resonant with the 5p$^2$P$_{3/2}$  (F$^\prime$) $\rightarrow$ nd$^2$D or 5p$^2$P$_{3/2}$ (F$^\prime$) $\rightarrow$ ns$^2$S transitions. 
The experimental setup is shown in Fig.~\ref{fig:setup}. A 780~nm probe beam and 479-486~nm coupling beam are counter propagated through a magnetically shielded Rubidium vapour cell at room temperature. Both beams initially have a radius of approximately 1~mm and are focused through the cell using a pair of lenses. The beams are focused in order to increase the coupling beam intensity, this allows EIT to be observed for transitions with small Einstein A-coefficients, such as transitions to Rydberg states. The coupling light is generated by a commercial high power frequency doubled laser (Toptica, TA-SHG). The probe light is provided by a commercial extended cavity diode laser (ECDL, Toptica DL-PRO ) which is frequency stabilized to the $^{87}$Rb 5s$^2$S$_{1/2}$ (F=2)   $\rightarrow$ 5p$^2$P$_{3/2}$  (F$^\prime$ = 3) resonance using FM transfer spectroscopy \cite{mcca08}. The probe beam is frequency modulated using an electro-optic modulator (EOM) at $\omega_{\rm m}/2\pi = 10$ MHz. The EOM is driven by an amplified (Mini-Circuits ZHL-3A) sinusoidal signal from an oscillator. Following the vapour cell the probe beam is incident on a fast photodiode (Hamamatsu APD C5460 10 MHz). The photodiode signal is frequency mixed (Mini-Circuits ZAD-6+) with the oscillator after passing through a phase shifter. The error signal is fed back to both the laser current and the external cavity piezo (using a Toptica FALC module with a 1~MHz cut-off filter). 

\begin{figure}[t] 

\includegraphics[]{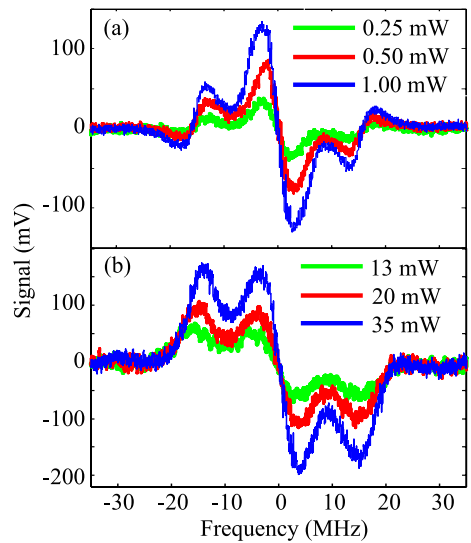}
\caption{ Locking signal at varying coupling beam powers. The probe beam was fixed at a power of 4 $\mu$W and was locked to the $^{87}$Rb 5s$^2$S$_{1/2}$ (F=2)   $\rightarrow$ 5p$^2$P$_{3/2}$  (F$^\prime$ = 3) transition  and modulated at 10 MHz using an electro-optic modulator (EOM). The coupling beam is tuned to the (a) 26D and (b) 70S Rydberg state and scanned through resonance.}
\label{fig:error_EIT}
\end{figure}

Examples of the resulting error signals while scanning the coupling laser are shown in Fig.~3. The laser system has been locked to Rydberg states between 19 and 57D and error signals observed up to 70S. For Rydberg states with intermediate values of the principal quantum number, e.g.\ $n=26$ in Fig.~\ref{fig:error_EIT}~(a), an error signal with good signal to noise is obtained with coupling laser powers of under 1~mW. For higher Rydberg states and transitions with smaller dipole matrix elements (the Einstein A-coefficient for the 5p-ns transitions is an order of magnitude less than 5p-nd) significantly higher powers are required, see Fig.~\ref{fig:error_EIT}~(b).

Although a direct beat measurement of the coupling laser was not feasible we use two approximate methods to evaluate the performance of the lock. Firstly we estimate a lower limit on the linewidth of the coupling laser system by measuring the rms noise of the error signal. As the amplitude of the error signal depends on the coupling laser power we can increase the gradient of the error signal without increasing the amplitude noise due to probe laser intensity fluctuations or electronic noise. This allows us to distinguish between frequency noise and amplitude noise and consequently we can determine what is limiting the stability of the lock. The rms noise of the locking signal (averaged over a period of 1 minute) when the laser is locked was measured and divided by the gradient of the unlocked signal to give a linewidth estimate for the 26D Rydberg state error signals. With a coupling power of 0.5~mW the gradient of the locking signal was 20~mV/MHz giving a linewidth of 200~kHz. A coupling power of 1.0~mW gave a gradient of 70~mV/MHz and a linewidth of 50~kHz. Finally, a power of 2.4~mW gave a gradient of 110~mV/MHz and a linewidth of 35~kHz, illustrating that for these parameters the linewidth is primarily limited by the gradient of the error signal and hence the coupling laser power used in the locking scheme. 

\begin{figure}[] 

\includegraphics[]{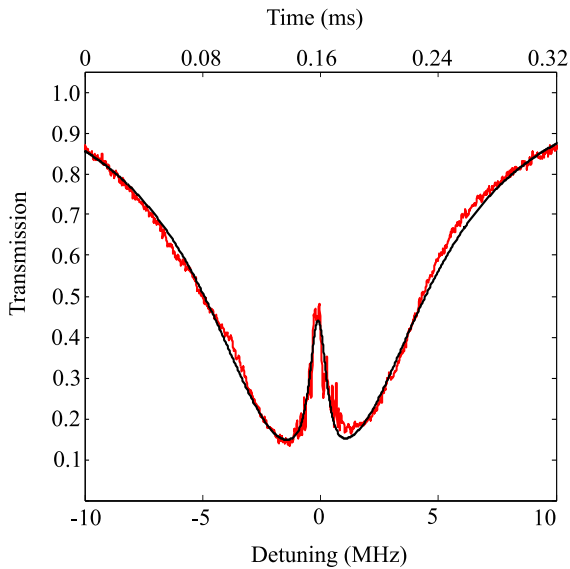}
\caption{Transmission of a 200 nW probe beam (red) as a function of probe beam detuning for the 5S$_{1/2}\rightarrow$5P$_{3/2}\rightarrow$26D transition. The probe laser is scanned over 20 MHz in 0.32 ms. The coupling beam power is 84 mW. The black line is the theoretical line of best fit from the model detailed in reference \cite{weat08} giving a combined laser linewidth of 280 $\pm$ 50 kHz.}
\label{fig:cold_atoms_EIT}
\end{figure}

Secondly, we obtain an estimate of the combined linewidth of the probe and coupling laser by performing  a two photon Rydberg excitation measurement on a cold Rydberg atom ensemble \cite{weat08}. The probe and pump beam were counterpropagated through a cloud of laser cooled Rb atoms. The probe beam is scanned through the $^{85}$Rb D$_2$ F = 3 $\rightarrow$ F$^\prime$ = 4 transition using an acousto-optic modulator (AOM) in a time of 0.32~ms to produce the EIT spectra in Fig.~\ref{fig:cold_atoms_EIT}. The theoretical model used to fit the data indicates a combined relative laser linewidth of 280 $\pm$ 50~kHz. The transit time broadening gives a contribution of less than 40~kHz.

Although the spectroscopy experiment gives a practical indication of the performance of the stabilized laser it underestimates the longer term linewidth as the measurement was performed over a short time scale of less than 100~$\mu$s. The linewidth of the probe laser was evaluated as the stability of the coupling laser is dependent on the stability of probe. To give an indication of the longer term linewidth we performed a beat measurement between the probe laser and two other 780~nm lasers which were locked using an equivalent method. By averaging over different numbers of measurements the linewidth was determined over time scales between 3.5 seconds and 170 seconds. The linewidth of the probe laser was found to vary between approximately 200 and 350~kHz which is consistent with the cold atom spectroscopy measurement.

In summary, we have demonstrated a laser frequency stabilization to excited state transitions using cascade EIT. Our case of Rydberg excitation displays the power of the method as the Einstein-A coefficients to Rydberg states are many orders of magnitude smaller than those of the Rb D lines. We expect this approach can be applied to many different excitation schemes in atomic and molecular physics experiments. 

After submission we became aware of a more recent preprint on two-color modulation transfer spectroscopy \cite{Oroz08}, which demonstrated signals associated with excited to excited transitions based on both absorption (similar to Ref. 19) and frequency modulation (as in our work).

We are grateful to I. G. Hughes and S. L. Cornish for the loan of equipment. We thank the EPSRC for financial support.


\begin{thebibliography}{99}

\bibitem{hall06}
J. L. Hall,
Rev. Mod. Phys. {\bf78}, 1279 (2006).

\bibitem{cher94}
B. Cher{\'o}n, H. Gilles, J. Hamel, D. Morau, and H. Sorel 
J. Physique III, {\bf 4}, 401 (1994).

\bibitem{corw98}
K. L. Corwin, Z.-T. Lu, C. F. Hand, R. J. Epstein and C. E. Wieman,
Appl. Opt. {\bf 37}, 3295 (1998).


\bibitem{mill07}
A. Millett--Sikking, I. G. Hughes, P. Tierney, and S. L. Cornish,
J. Phys. B. {\bf 40}, 187 (2007).


\bibitem{harr08}
M. L. Harris, S. L. Cornish, A. Tripathi, and I. G. Hughes,
J. Phys. B. {\bf 41}, 085401 (2008).


\bibitem{tino03}
T. Petelski, M. Fattori, G. Lamporesi, J Stuhler and G. M. Tino,
Eur. Phys. J. D {\bf22} 279 (2003).


\bibitem{wasi02}
G. Wasik, W Gawlik, J. Zachorowski  and W. Zawadzki
Appl. Phys. B {\bf 75} 613 (2002).

\bibitem{pear02}
C. P. Pearman, C. S. Adams, S. G. Cox, P. F. Griffin, D. A. Smith, and I. G. Hughes,
J. Phys. B {\bf 35}, 5141 (2002).

\bibitem{robi02}
N. P. Robins, B. J. J. Slagmolen, D. A. Shaddock, J. D. Close, and M. B. Gray,
Opt. Lett. {\bf 27}, 1905 (2002).

\bibitem{jund03}
G. Jundt, G.T. Purves, C.S. Adams, and I.G. Hughes,
Eur. Phys. J. D. {\bf 27}, 273 (2003).

\bibitem{bjor80}
G. C. Bjorklund, Opt. Lett. {\bf 5}, 15 (1980).

\bibitem{shir82}
J. H. Shirley, Opt. Lett. {\bf 7}, 537 (1982).

\bibitem{zhan03}
J. Zhang, D. Wei, C. Xie, and K. Peng,
Opt. Express. {\bf 11}, 1338 (2003).

\bibitem{mcca08}
D. J. McCarron, S. A. King and S. L. Cornish
Meas. Sci. Technol. {\bf19} 105601 (2008).

\bibitem{harris}
K.-J. Boller, A. Imamo\u{g}lu, and S. E. Harris,
Phys. Rev. Lett. {\bf 66}, 2593 (1991).

\bibitem{flei05}
M. Fleischhauer, A. Imamo\u{g}lu and J. P. Marangos,
Rev. Mod. Phys. {\bf 77}, 633 (2005).


\bibitem{bell07}
S. C. Bell, D. M. Heywood, J. D. White, J. D. Close, and R. E. Scholten,
Appl. Phys. Lett. {\bf 90}, 171120 (2007).

\bibitem{scho08}
D. V. Sheludko, S. C. Bell, R. Anderson, C. S. Hofmann, E. J. D. Vredenbregt, and R. E. Scholten
Phys. Rev. A {\bf 77}, 033401 (2008).

\bibitem{bohl06}
P. Bohlouli-Zanjani, K. Afrousheh, and J. D. D. Martin,
Rev. Sci. Instrum. {\bf 77}, 093105 (2006).



\bibitem{gallagher} T. F. Gallagher, {\it Rydberg atoms} (Cambridge University Press, Cambridge 1994).



\bibitem{moha07}
A. K. Mohapatra, T. R. Jackson, and C. S. Adams, Phys. Rev. Lett. {\bf 98}, 113003 (2007).

\bibitem{baso08}
M. G. Bason, A. K. Mohapatra, K. J. Weatherill, and C. S. Adams,
 Phys. Rev. A. {\bf 77}, 032305 (2008).


\bibitem{weat08}
K. J. Weatherill, J. D. Pritchard, R. P. Abel, M. G. Bason, A. K. Mohapatra, and C. S. Adams.
J. Phys. B. {\bf41}, 201002 (2008).

\bibitem{moha08}
A. K. Mohapatra, M. G. Bason, B. Butscher, K. J. Weatherill, and C. S. Adams,
Nature Phys. {\bf4}, 890  (2008).




\bibitem{bjor83}
G. C. Bjorklund, M. D. Levenson, W. Lenth, and C. Ortiz,
Appl. Phys. B. {\bf 32}, 145 (1983).

\bibitem{hall83}
J. L. Hall, F. V. Kowalski, J. Hough, G. M. Ford, A. J. Munley, and H. Ward,
Appl. Phys. B. {\bf 31} 97 (1983).

\bibitem{Oroz08}
A. P\'{e}rez Galv\'{a}n, D. Sheng, L. A. Orozco and Y. Zhao,
Can. J. Phys. {\bf87}(1):95-100 (2009).




\end{thebibliography}
\end{document}